\documentstyle [prb,aps,preprint] {revtex}

\begin{document}

\draft
\preprint{VA70/99-1}
\title{
All-electron and pseudopotential study of the spin-polarization
of the V(001) surface: LDA versus GGA.} 

\author{R. Robles, J. Izquierdo, A. Vega and L. C. Balb\'as}
\address{Departamento de F\a'{\i}sica Te\a'orica, At\a'omica, Molecular y
Nuclear\\
Universidad de Valladolid, E-47011 Valladolid, Spain}

\date{\today}
\maketitle
\begin{abstract}
The spin-polarization at the V(001) surface has been studied
by using different local (LSDA) and semilocal (GGA) approximations
to the exchange-correlation potential of DFT 
within two {\it ab initio} methods:
the all-electron TB-LMTO-ASA and the pseudopotential LCAO code
SIESTA (Spanish Initiative for Electronic Simulations with
Thousands of Atoms).
A comparative analysis is performed first for the bulk and then for
a N-layer V(001) film (7$\le$N$\le$15).
The LSDA approximation leads to a non magnetic V(001) surface
with both theoretical models in agreement (disagreement) with magneto-optical Kerr
(electron-capture spectroscopy) experiments.
The GGA within the pseudopotential method needs thicker slabs than the LSDA to yield zero moment at the
central layer,
giving a high surface magnetization (1.70 Bohr magnetons),
in contrast with the non magnetic solution obtained
by means of the all-electron code.
\end{abstract}

\section {Introduction}

The understanding and prediction of the magnetic behavior of vanadium-based 
systems
has motivated numerous experiments and calculations over the last fifteen
years. Vanadium is one of those paramagnetic metals \cite{Alik77}
that can exhibit magnetism
under certain conditions (loss of coordination, hybridization with a
ferromagnet), due to its large paramagnetic susceptibility
\cite{janak77,Hattox73}. 
For instance, Akoh and Tasaki \cite{Akoh77} have reported 
large localized magnetic moments in hyperfine particles of V, and
several experimental groups have demonstrated the existence of an induced
magnetization at the V interface 
for V overlayers on
Fe substrates
\cite{Walker94,Fuchs96,Finazzi98}
and Fe/V multilayers
\cite{Harp95,Tomaz97,Schwickert98}.
Although these trends are clear, in other
aspects there has not been consensus.
One controversial aspect concerns the short or long range induced
spin-polarization in V. Two of us have reported on this problem in
a recent work \cite{Izquierdo99} where the reader can find a
complete review.
Another unclear aspect, that gave rise to an interesting discussion
more than ten years ago, was the magnetic character of the V(001) surface.
Rau {\it et al} \cite{Rau} through electron-capture spectroscopy
concluded the existence of ferromagnetic order at the V(001) p(1x1) 
surface and
on the V monolayer supported on Ag(001), whereas Fink {\it et al}
\cite{Fink90} through magneto-optical Kerr measurements did not
find magnetization in ultrathin epitaxial films of V on Ag substrates.
{\it Ab initio} full-potential linearized augmented plane wave
(FLAPW) calculations within the density functional framework
using the von Barth-Hedin \cite{Barth72} local spin density approximation 
(LSDA) for the exchange and correlation  (XC) potential
found no surface magnetization \cite{Ohnishi85}, which was since then
admitted, and corroborated using other all-electron
methods with LSDA (LMTO-ASA)
\cite{Turek98}.

Very recently, Bryk {\it et al} \cite{Bryk00} have reported a theoretical
work entitled "Magnetization of the V(001) surface 
in the generalized gradient approximation".
They performed an {\it ab initio} calculation for a
seven layer V(001) film using the plane-waves method with
the ultrasoft (US) non-normconserving pseudopotential (PP) of Vanderbilt
\cite{Vanderbilt90}, 
and the generalized gradient approximation (GGA)
of Perdew \cite{Perdew91} (PW91) for the XC potential,
 obtaining a magnetic moment of 1.70$\mu_B$ at the unrelaxed surface
plane ( 1.45$\mu_B$ if relaxed).
These authors believe that their prediction of a magnetic V(001) surface is
more likely to be correct than the contrary result obtained earlier 
\cite{Ohnishi85} from 
local spin density approximation (LSDA), and consequently they request 
new measurements to test the prediction. 
Besides, although they correctly find the paramagnetic bulk ground state 
for V at the calculated equilibrium lattice constant,
they also obtain, at the experimental lattice constant,
the ferromagnetic phase, with
0.15$\mu_B$ per atom, lying only 0.7 meV below the paramagnetic 
one. In the opinion of the authors, this energy is small compared with the
difference between the calculated magnetic and non magnetic surface energies,
concluding that the GGA overestimates the magnetization of the interior
layers, but does not alter their prediction of magnetic V(001) surface.
An even more recent calculation of the magnetic structure of V(001) surface using the {\it all-electron} FLAPW method
and the PW91 form of GGA \cite{Perdew91}, by Bihlmayer, Asada and Bl\"ugel
\cite{Bihlmayer00}, conclude that ``in very thin V films a surface magnetic
moment can be stabilized, while for thicker and relaxed films
no surface magnetism can be found".
 
The suspicion that the PW91 form of GGA might incorrectly tip
 metals with large LSD spin-susceptibility enhancements, like Pd and V,
into the ferromagnetic state, was posed in the first paper testing the
PW91 functional \cite{Perdew92}.
Subsequent calculations using all-electron
\cite{Singh92,Ozolins93,Holzwarth97,Kresse99} and/or
pseudopotentials \cite{Ozolins93,Holzwarth97,Kresse99,Izquierdo00} methods,
coincide pointing out the trend of GGA's to enhance magnetism of
magnetic materials and susceptibilities of non magnetic materials
when compared with results from analogous LSDA calculations.
Singh and
Ashkenazi \cite{Singh92}
using the all-electron FLAPW method confirmed that bulk V and Pd
are correctly predicted as paramagnetic by the PW91 GGA approximation,
but conclude also that GGA's do not have greater precision
than the LSDA for studying transition metals (TM)
and specially for magnetic materials.
In Ref. [20] it is also noted that there are considerable differences
between the predictions of the GGA's from Langreth-Mehl-Hu (LMH)
\cite{Langreth81} and PW91, which gives room for more consistent
improvements on the LSDA.
Ozolins and K\"orling \cite{Ozolins93}, using the
full-potential linear muffin-tin-orbital (FP-LMTO) method
\cite{Andersen84,Andersen86} found that the
PW91 GGA predicts more accurately than the LSDA the
structural properties of non magnetic 3$d$, 4$d$ and 5$d$
TM.
Also in Ref. [21] it is suggested that
a full-potential treatment, instead of the atomic-sphere
approximation (ASA), is important for GGA calculations,
because the inaccuracies introduced by the ASA are of the
same magnitude as the gradient corrections.
Apparently, an ASA GGA calculation average out the angular parts of
the GGA and thus is missing what could be a portion of the
GGA contribution.
Norm conserving and ultrasoft pseudopotentials has been compared for
first-row and transition elements by Kresse and Hafner
\cite{Kresse94} from the point of view o numerical
performance achieving good convergence and
transferability properties.
The extreme softening of non-normconserving Vanderbilt pseudopotentials
without loss of accuracy compared to very hard and accurate pseudopotentials
(working at much more higher cutoff energy) has been demonstrated
recently by Furthm\"uller and coworkers
\cite{Furthmuller00}.
On the other hand, the use of pseudopotentials, particularly
the US-PP of Vanderbilt
\cite{Vanderbilt90}, has been tested by Holzwarth et al
\cite{Holzwarth97} and by Kresse and Joubert
\cite{Kresse99} against all-electron FLAPW and
projector augmented wave (PAW) \cite{Blochl94} results for structural
properties of several metals, including vanadium
\cite{Holzwarth97,Kresse99}, and the magnetic properties
of Fe, Co and Ni \cite{Kresse99}. Whereas
Holzwarth et al \cite{Holzwarth97} conclude that the
structural properties of bcc V are represented equally well
by the PAW, LAPW and PP methods,
Kresse and Joubert \cite{Kresse99} point out that for
TM at the left side of the Periodic Table it is very desirable
to include semicore states as valence states in the
pseudopotentials.
The LSDA (GGA) magnetic moments of bcc Fe calculated
in Ref. [23] are slightly (considerably) larger
within the PP method than within the AE methods FLAPW and PAW.
On the other hand, in both PP and AE calculations, the
GGA magnetic moments of Fe, Co and Ni are systematically larger than the
LSDA ones.
The same trends are obtained in Ref. [24]
by comparing
calculations using the PP-LCAO code SIESTA
\cite{Izquierdo00}
with available AE results for the magnetic
moment of bulk bcc Fe, the (001) Fe surface, and small
clusters of Fe on the (001) Ag surface.

The work of Bryk {\it et al} \cite{Bryk00} opens again clearly the 
discussion about the magnetic
character of V(001) and besides, from the theoretical point of view,
it offers a new benchmark to test different {\it ab initio} methods and different
XC functionals.
Therefore, the need of further calculations is evident. 
In this work, calculations have been done 
using two different {\it ab initio} methods based on the density functional 
theory (DFT) \cite{Hohenberg64}, namely,
$i$) AE tight-binding linear muffin
tin orbitals (TB-LMTO) method, in the 
scalar relativistic version and the 
ASA \cite{Andersen84,Andersen86}, and $ii$)  the 
SIESTA LCAO method \cite{siesta99}, using the Troullier-Martins
pseudopotentials \cite{Troullier91}. 
For the XC functional,
we have used within TB-LMTO four different approximations: 
two LSDA versions -the
Hedin-Barth \cite{Barth72} and the Vosko-Wilk-Nusair \cite{Vosko80}-, 
and two GGAs -the PW91 \cite{Perdew91} and the
Langreth-Mehl-Hu (LMH) \cite{Langreth81}-. 
The XC functionals used within the pseudopotential code SIESTA 
\cite{siesta99} 
are the Perdew-Zunger (PZ) parameterization for the LSDA \cite{Perdew81} and the
Perdew-Burke-Ernzerhof (PBE) \cite{Perdew96} form of GGA.
The soft ionic pseudopotential used in SIESTA
is generated according to the procedure of
Troullier and Martins\cite{Troullier91} from the atomic
configurations [Ar]3d$^3$4s$^2$ for V 
with core
radii for the the s, p and d components of 2.35, 2.70 and 2.35
a.u. respectively. 
The partial core-correction
for non-linear XC \cite{Louie82} has been included. A careful study of the
optimum core-correction radius leads to a value of
0.8 a.u.   

Let us first discuss the bulk bcc case. 
The GGA's equilibrium lattice constants are,
within the TB-LMTO,
(in a.u.) 5.68 (5.79)  
from the LMH (PW91) calculations, and 5.71 for the SIESTA-PBE
calculation. 
The LSDA's equilibrium lattice constant is 5.61 a.u. for the different
LSDA approximations in both 
TB-LMTO and SIESTA calculations.
With both AE and PP methods, and for all the XC-approximations
used we obtain a paramagnetic
ground state at the experimental lattice constant (5.73 a.u.), 
in agreement with the experimental findings \cite{Alik77}.
This  contrasts with the GGA result of Bryk {\it et al} \cite{Bryk00}
where bulk V at the experimental 
lattice constant is ferromagnetic (F) with 0.15$\mu_B$ per atom,
although they find the paramagnetic state at the equilibrium
lattice constant in their calculations.
It is a general trend that when expanding the lattice, the electronic 
localization
increases and the kinetic energy of the system decreases so that
spin-polarization is favored. It is therefore expected to find a magnetic
phase transition in V bulk in this context. 
Moruzzi and Marcus
\cite{Moruzzi90} studied ten years ago this magnetic transition
in V bulk through total-energy band calculations using
the LSDA in the augmented-spherical-wave (ASW) method.
They found that the paramagnetic solution is the ground state of
V bulk for the experimental lattice constant and for lattice expansions 
of less than 12\%. For this expansion they find a transition to the
antiferromagnetic (AF) solution  
(always more stable than a low-spin F solution).
Earlier spin-polarized augmented plane wave (APW) calculations for
bcc vanadium \cite{Hattox73} found also that a non magnetic to ferromagnetic
transition occurs abruptly for a lattice constant about 25\% 
larger than the equilibrium value.
We have also tested volume expansions and within all XC-functionals 
considered in our
TB-LMTO and SIESTA calculations, the paramagnetic ground state 
persists at least up to an expansion of 10\%.

In the case of the seven layer V(001) film at the bulk experimental lattice
constant, the LSDA gives no surface magnetization in both our AE and PP calculations, 
in agreement with
previous first principles LSDA results \cite{Ohnishi85,Turek98}. Concerning the GGA, our 
all electron results 
are reported in figure 1 for both the PW91 and LMH versions of GGA.
Here a surface magnetization is obtained, although it is much lower
than the value reported by Bryk {\it et al} \cite{Bryk00}
using pseudopotentials.
They obtain 1.70$\mu_B$ at the surface plane in contrast with
our values of 0.66$\mu_B$ for the PW and 0.25$\mu_B$ for
the LMH cases (Fig. 1). Notice also the large difference obtained 
with different versions of GGA, 
as already observed in Ref. [20], and that
for volume expansions of the order of 2\% that difference become 
negligeable (Fig. 1b).
We see then, that the AE
surface magnetization is still far from the PP value of 1.7$\mu_B$ obtained by
Bryk {\it et al} \cite{Bryk00}.
This system constitutes a good test for an
{\it ab initio} pseudopotential calculation, in order, to see whether
or not the pseudopotential code give rise to an overestimation of
the V magnetic moment compared with a typical AE method.
For this purpose we have repeated the same seven-layer calculations with SIESTA
and the PP of Troullier-Martins.
The LSDA again leads to the paramagnetic solution whereas the GGA
leads to a magnetic moment of 1.77$\mu_B$ which is similar to 
the value reported by Bryk {\it et al} \cite{Bryk00}.
This results indicate that the GGA within the pseudopotential calculations
enhances even more than within AE methods the V (001) magnetic moment.
We note, however, a difference between
our PP-GGA results and those of Bryk {\it et al} \cite{Bryk00} for
the seven layer V film. At the experimental lattice constant,
they achieve in the central layer 
the convergence of the magnetic moment to their bulk value (0.15$\mu_B$) 
and, consequently, they use the surface
layer to discuss the V(001) surface.
However, our results for the central layer 
do not converge to our paramagnetic bulk with
just seven layers. As, 
 we increase the number of layers in the film, 
the calculated magnetic moment at the center decreases. 
For 15 layers the local
moment at the center is zero as in the bulk and 
we can discuss then the V(001) surface magnetism in terms of
 the surface layer.
The interesting result is that when we consider more than seven layers in the slab the V(001) surface is non magnetic
with the all-electron TB-LMTO method, independently of the approximation used for the XC potential
(LDA or GGA), 
whereas with the pseudopotential code SIESTA, even for calculations with 15 layers of V the GGA,
still produces a high magnetic moment at the surface
(see table 1), comparable with that obtained by Bryk et al
\cite{Bryk00} (with LDA we obtain the non magnetic surface).

In resume our results indicate the well known tendency 
\cite{Singh92,Izquierdo00}
of GGA to produce larger magnetic moments than the LSDA, 
 but also that the convergence of GGA to the
bulk value at the central layer of a slab is slower than for LSDA.
Taking care of these convergence problems, 
the V(001) surface is non magnetic with both the LSDA and GGA 
within the all-electron TB-LMTO method, whereas the pseudopotential code
with GGA still produces a high magnetic moment.
It is interesting to note that our pseudopotentials and basis in SIESTA
are different from those of Bryk et al
\cite{Bryk00}.
On the other hand, in view of the controversial theoretical predictions,
new experiments concerning the magnetic character of the (001) surface
of vanadium are needed to confirm or to correct the
earlier findings of Rau and coworkers
\cite{Rau}.

\acknowledgments
We acknowledge financial support of DGICYT of Spain(Grant
PB98-0368-C02) and of Junta de Castilla y Le\'on (Grant VA70/99). J.I.
and R.R acknowledge the F.P.I grants from M.E.C. of Spain.

\begin{figure}
\caption{
Magnetic moments profile for the 7-layer V(001) film
obtained with TB-LMTO method and two GGA versions 
(LMH \cite{Langreth81} and PW91 \cite{Perdew91}). Upper 
panel corresponds to the experimental lattice
constant whereas lower panel corresponds to a lattice
expansion of 2\%.}
\label{Fig. 1}
\end{figure}

\narrowtext 
\begin{table}
\caption{
Local magnetic moment obtained at the surface of a 7-layer and 15-layer
V(001) film with the different versions of the GGA in
the all-electron TB-LMTO and pseudopotential
SIESTA methods for the experimental bulk lattice constant of vanadium. For slabs thicker that 7 layers the
TB-LMTO yields non magnetic ordering at the V(001) surface. 
}
\begin{tabular}{c|ccc}
    & TB-LMTO(LMH) & TB-LMTO(PW91) & SIESTA(PBE) \\ 
\hline
7 layers & 0.25 & 0.66 & 1.77 \\
15 layers & 0.00 & 0.00 & 1.70 \\
\end{tabular}
\end{table}   

\end {document}
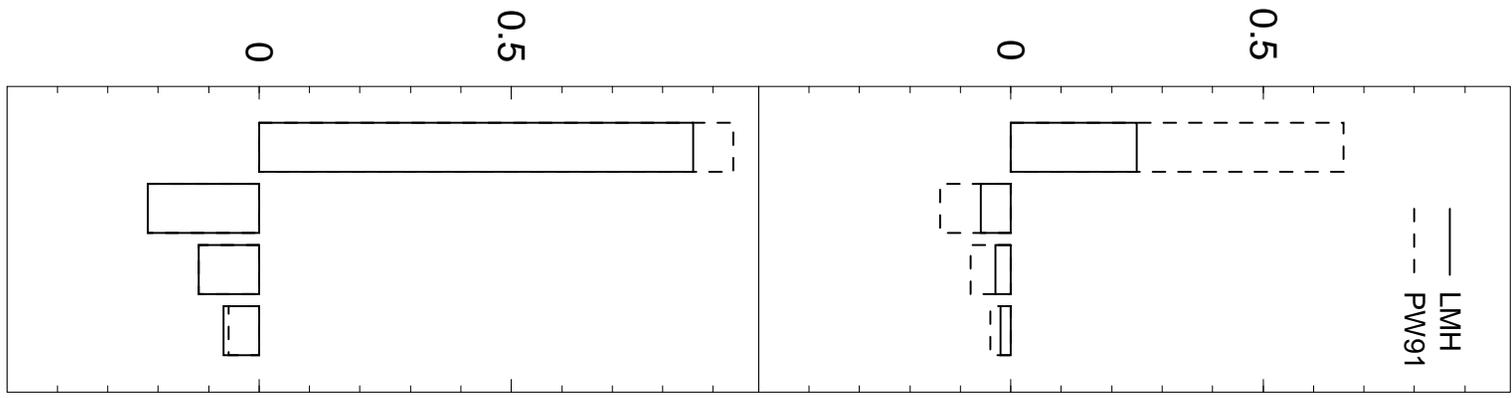
